\documentclass[a4pa per,11pt]{article} 
\usepackage[english]{babel}           
\usepackage[utf8]{inputenc}          
\usepackage{graphicx}    
\usepackage{color}      
\usepackage{anysize}     
\usepackage{multicol}    
\usepackage{bm}          
\usepackage{subcaption}
\usepackage{eurosym}     
\usepackage{amsthm}      
\usepackage{gensymb}
\usepackage{ amssymb }
\usepackage{float}
\usepackage{lineno}     
\usepackage{multirow}
\usepackage{dsfont}
\usepackage{float}
\usepackage[nottoc]{tocbibind}

\marginsize{2.5cm}{2.5cm}{1.5cm}{1.5cm} 
\usepackage{ amssymb }
\usepackage{ amsmath }
\usepackage{graphicx}
\usepackage{hyperref} 

\begin{document}
\begin{titlepage}
\vfill

\begin{center}

{\centering\LARGE\bf Fast simulation of a forward detector at 50 and 100 TeV proton-proton colliders}

	\vspace{2cm}

Veronika Chobanova, Diego Mart\'inez Santos, Claire Prouve, Marcos Romero Lamas

{\it Instituto Galego de F\'isica de Altas Enerx\'ias (IGFAE), Universidade de Santiago de Compostela, Santiago, Spain }

\end{center}
\vspace{2cm}

\begin{abstract}
We evaluate the performance of an LHCb-like detector using a fast simulation  of proton-proton collisions at center-of-mass energies of 50 and 100 TeV. The study shows that detector acceptances and resolutions could be similar to those at the LHC. Together with the increase of production cross-sections of light particles with higher energy, such a hypothetical experiment could reach unprecedented sensitivities in flavor observables. 
\end{abstract}
\end{titlepage}

\section{Introduction}
Hadron colliders with energies well above the LHC have been proposed. Notable examples are SppC~\cite{SPPC} and FCC-hh~\cite{FCC-hh}. In this paper, we evaluate the acceptances and resolutions that a forward detector similar to the LHCb Upgrade would have in those conditions. In Section~\ref{sec:gen} we discuss the proton-proton collision simulation. In Section~\ref{sec:det} we discuss
the detector simulation. In Sections~\ref{sec:effs} and~\ref{sec:res} we discuss the efficiencies and resolutions obtained in our simulation for the different center-of-mass energies of the proton-proton collisions. Conclusions are drawn in Section~\ref{sec:conclusions}.

\section{Sample generation}
\label{sec:gen}

Proton-proton collisions are simulated with \textsc{Pythia 8.1}~\cite{pythia} at center-of-mass energies of 13, 50, and 100 TeV where, as expected, the higher the energy, the larger the production cross-sections. This increase, as obtained with \textsc{Pythia}, is shown in Table~\ref{tab:xsections}. The table also shows the anti-particle to particle production ratio. The four-momenta of the particles of interest that have been produced by \textsc{Pythia} are stored in a file for further processing.

\begin{figure}[H]
    \centering
    \includegraphics[width=0.45\textwidth, page=1]{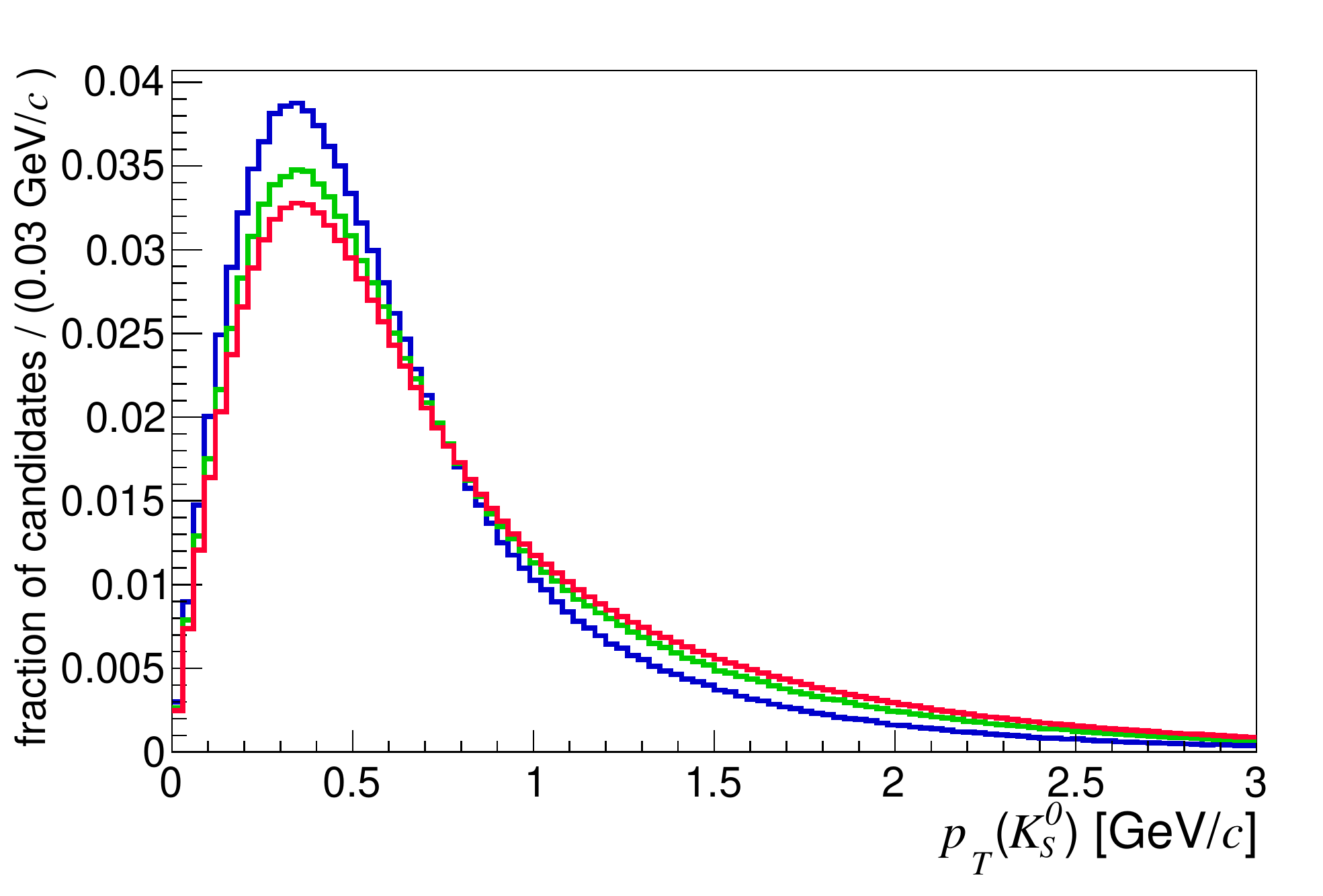}
    \includegraphics[width=0.45\textwidth, page=1]{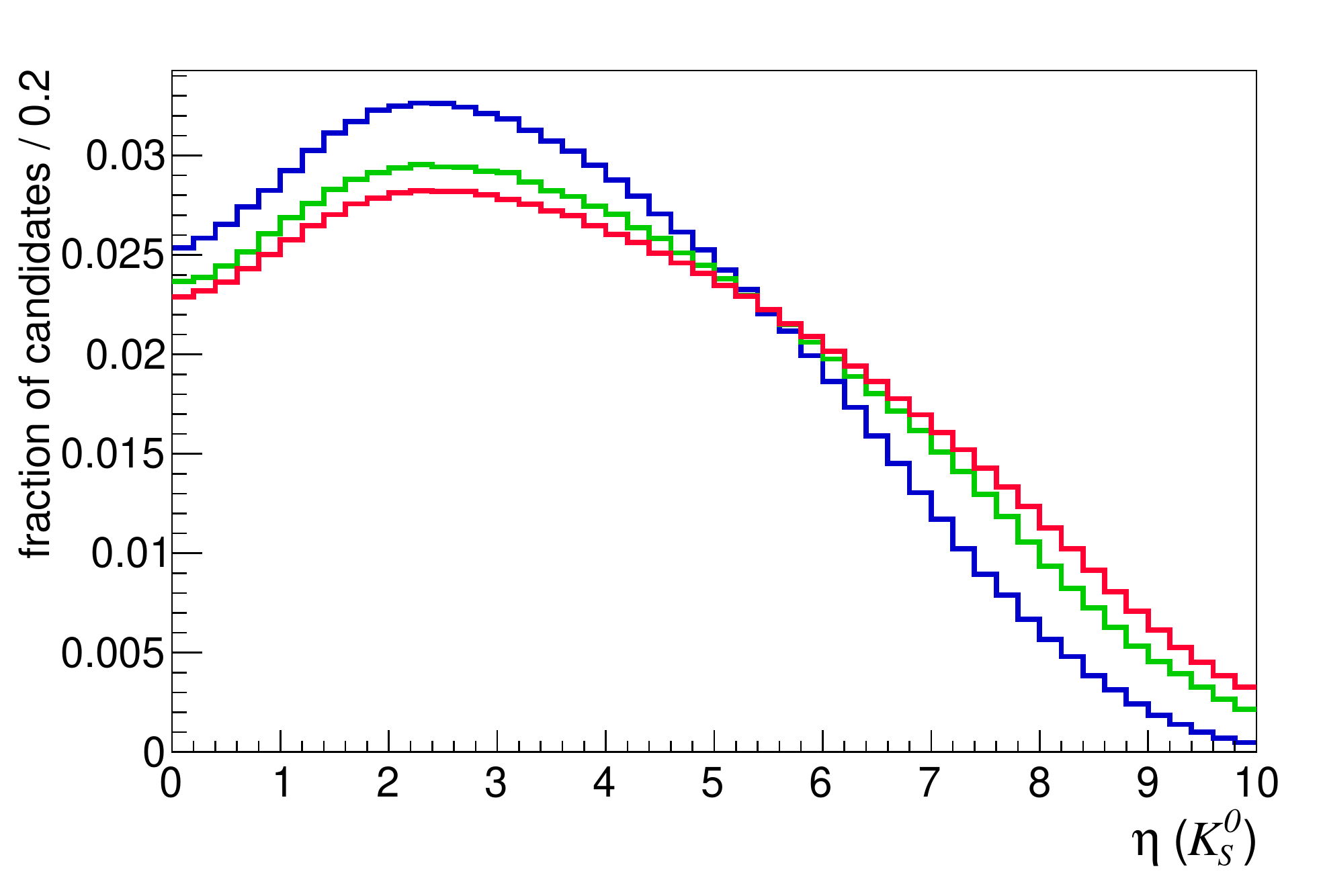}
    \centering
    \includegraphics[width=0.45\textwidth, page=2]{transverseMomentumMothers.pdf}
    \includegraphics[width=0.45\textwidth, page=2]{pseudorapidityMothers.pdf}
    \centering
    \includegraphics[width=0.45\textwidth, page=3]{transverseMomentumMothers.pdf}
    \includegraphics[width=0.45\textwidth, page=3]{pseudorapidityMothers.pdf}
    \caption{Left: Transverse momentum of parent particles produced in our simulated collisions at different energies. Right: Pseudorapidity of the same particles . Blue: 13 TeV. Green: 50 TeV. Red: 100 TeV.}
    \label{fig:kinematics}
\end{figure}

\begin{figure}
    \centering
    \includegraphics[width=0.45\textwidth, page=1]{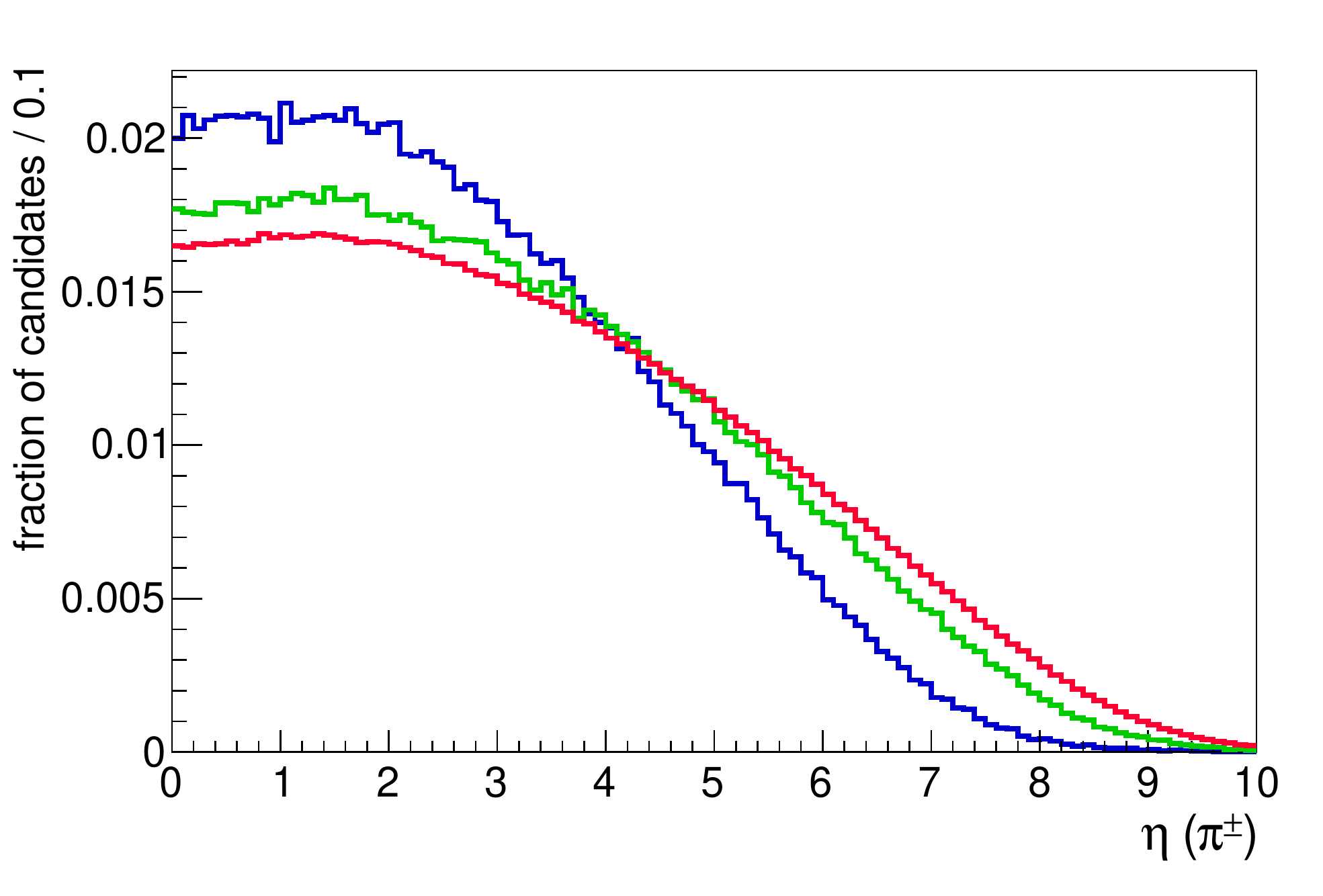}
    \includegraphics[width=0.45\textwidth, page=2]{pseudorapidityDaughters.pdf}
    \caption{Left: Pseudorapidity of the pions from $D^0\rightarrow \pi^+\pi^-\mu^+\mu^-$ decays produced in our simulated collisions at different energies. Right: Pseudorapidity of the muons from $\tau^-\rightarrow\mu^-\mu^+\mu^-$ decays produced in our simulated collisions at different energies. The LHCb acceptance is highest at a pseudorapidity between 3 and 5. Blue: 13 TeV. Green: 50 TeV. Red: 100 TeV.}
    \label{fig:kinematicsdaughters}
\end{figure}

\begin{table}[htb]
\centering
\caption{Production cross-sections for different particle species at 50 TeV ($\sigma_{P}^{50}$) and 100 TeV ($\sigma_{P}^{100}$) normalized to 13 TeV ($\sigma_{P}^{13}$). The particle/anti-particle cross-section ratios are also provided ($\sigma_P/\sigma_{\bar{P}}$). The numbers are obtained using \textsc{Pythia 8.1} } 
\begin{tabular}{l|l|l|l|l|l}
Particle & $\sigma_P^{50}/\sigma_{P}^{13}$ &  $\sigma_P^{100}/\sigma_{P}^{13}$  &  $\sigma_{\bar{P}}^{13}/\sigma_P^{13}$ &  $\sigma_{\bar{P}}^{50}/\sigma_P^{50}$ &  $\sigma_{\bar{P}}^{100}/\sigma_P^{100}$ \\ \hline

$\tau^-$ & 2.5 & 3.8 & 1.0 & 1.0 & 0.99 \\
$B_c^+$ & 7.2 & 14 & 1.6 & 0.97 & 0.98 \\
$B_d^0$ & 3.1 & 5.1 & 0.99 & 0.99 & 0.99 \\
$B^+$ & 3.1 & 5.1 & 0.99 & 1.0 & 1.0 \\
$B_s^0$ & 3.2 & 5.4 & 1.0 & 1.0 & 0.99\\
$\Lambda_b^0$ & 3.0 & 4.1 & 0.89 & 0.89 & 0.90  \\
$K^0$ & 1.6 & 2.0 & 0.98 & 0.99 & 0.99\\
$K^+$ & 1.6 & 2.0 & 0.97 & 0.98 & 0.99 \\
$\Lambda^0$ & 1.5 & 1.9 & 0.86 & 0.90 & 0.92 \\
$\Sigma^+$ & 1.5 & 1.9 & 0.89 & 0.93 & 0.94 \\
$\Omega^-$ & 1.6 & 2.1 & 0.98 & 0.95 & 0.96 \\
$\chi^-$ & 1.6 & 2.0 & 0.96 & 0.97 & 0.97 \\
$\chi^0$ & 1.6 & 2.0 & 0.95 & 0.96 & 0.97 \\
$D^{0,+}$ & 2.2 & 3.1 & 1.0 & 1.0 & 1.0 \\
$\Lambda_c^+$ & 2.1 & 2.9 & 0.90 & 0.93 & 0.94 \\

\end{tabular}
\label{tab:xsections}
\end{table}
 \clearpage
\setlength{\parindent}{0pt}
Subsequently, parent particles are decayed following either phase space or custom \texttt{EvtGen}~\cite{Lange:2001uf} implementations for more detailed decay models (such as $B_s^0\rightarrow J/\psi(\mu\mu)K^+K^-$). No bremsstrahlung is included in the generation of the decays. The procedure is performed for center-of-mass energies of 13, 50 and 100 TeV. The generated samples are listed in Table~\ref{tab:gen}\footnote{We use {\it DS}, for {\it Dark Sector} subscript to avoid confusion between the Dark Sector $\psi$ particle and the charmonia states}.
\\
\\
Figure~\ref{fig:kinematics} shows the transverse momentum distributions and the pseudorapidty distributions of $K_S^0$, $D^0$ and $B_s^0$ mesons directly produced in our simulated collisions. Figure~\ref{fig:kinematicsdaughters} shows the pseudorapity of daughter particles from our simulated $D^0\rightarrow \pi^+\pi^-\mu^+\mu^-$ and $\tau^-\rightarrow\mu^-\mu^+\mu^-$ decays. The LHCb acceptance is highest at a pseudorapidity between 3 and 5.

\begin{table}[]
\centering
\caption{Simulated samples for different energies. The parent particle produced by \textsc{Pythia} is requested to be within 400 mrad of LHCb's acceptance before proceeding to simulation.The $\Lambda_b^0$ decays correspond to dark matter searches linked to matter-antimatter asymmetry~\cite{Elor:2018twp}, where $\psi_{(DS)}$ decays to invisible particles, leading to missing transverse energy.}
\begin{tabular}{c|c|c|c}
Decay     &  13 TeV & 50 TeV & 100 TeV \\ \hline
$B_d^0\rightarrow\mu^+\mu^-$     &  41 k & 135 k & 228 k \\\
$B_s^0\rightarrow J/\psi(\mu\mu)K^+K^-$ & 17 k & 30 k & 30 k \\
$\Lambda_b^0\rightarrow K^+\pi^-\psi_{(DS)}(4340)$ & 4k & 12k& 20k \\
$\Lambda_b^0\rightarrow K^+\pi^-\psi_{(DS)}(1460)$ & 4k & 12k& 20k \\

$D^0\rightarrow K^+K^-$ & 138 k & 247 k & 3.5 M\\
$D^0\rightarrow K_S^0 \pi^+ \pi^-$ & 138 k & 247 k & 3.5 M\\
$D^0\rightarrow \pi^+\pi^-\mu^+\mu^-$ & 138 k & 247 k & 3.5 M\\
$K_S^0\rightarrow\mu^+\mu^- $ & 74 M & 100 M & 100 M \\
$\tau^-\rightarrow\mu^-\mu^+\mu^-$ & 2 k & 4 k & 67 k \\
\end{tabular}
\label{tab:gen}
\end{table}

\section{Detector simulation}
\label{sec:det}
\begin{figure}[H]
    \centering
    \includegraphics[width=0.45\textwidth]{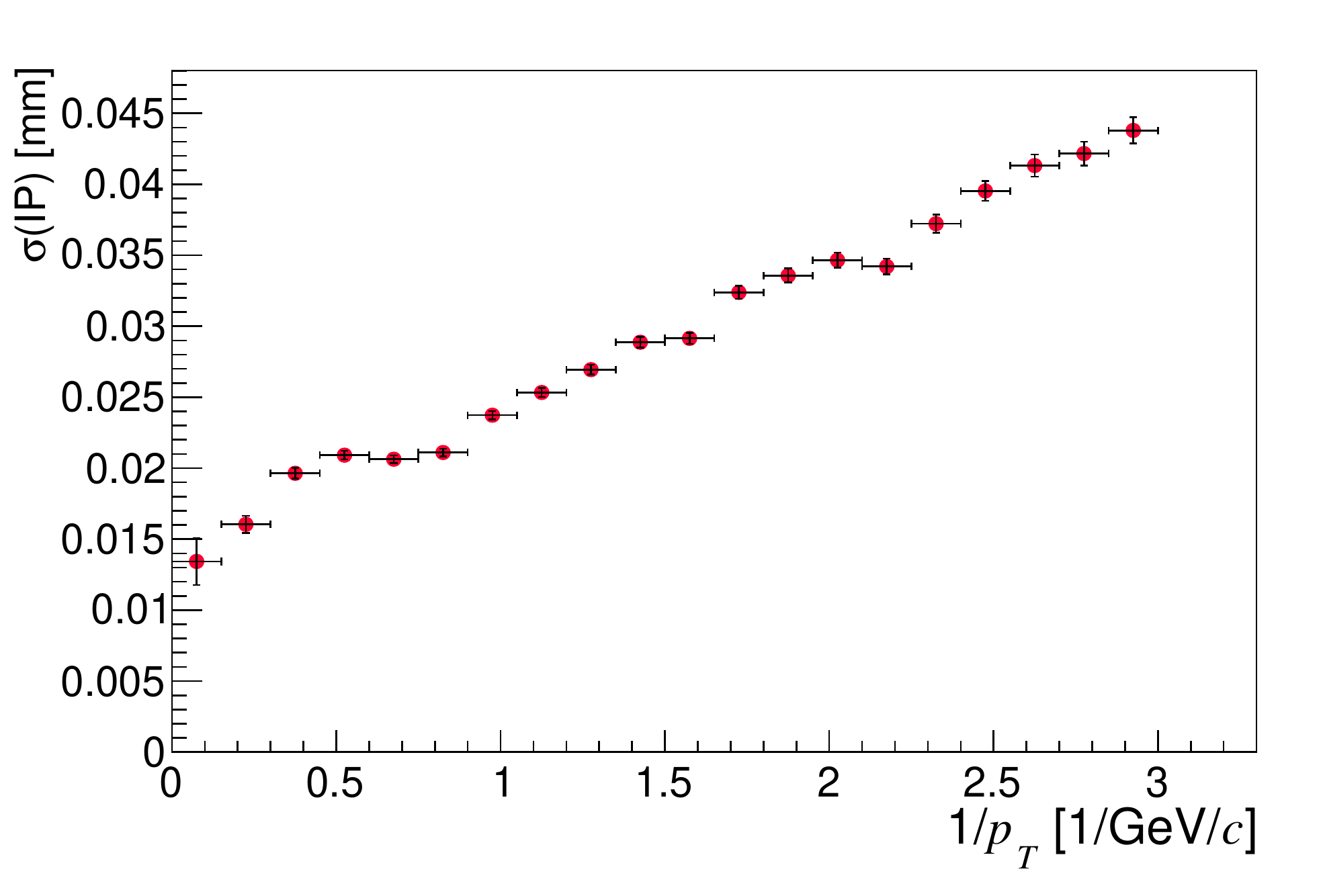}
    \includegraphics[width=0.45\textwidth]{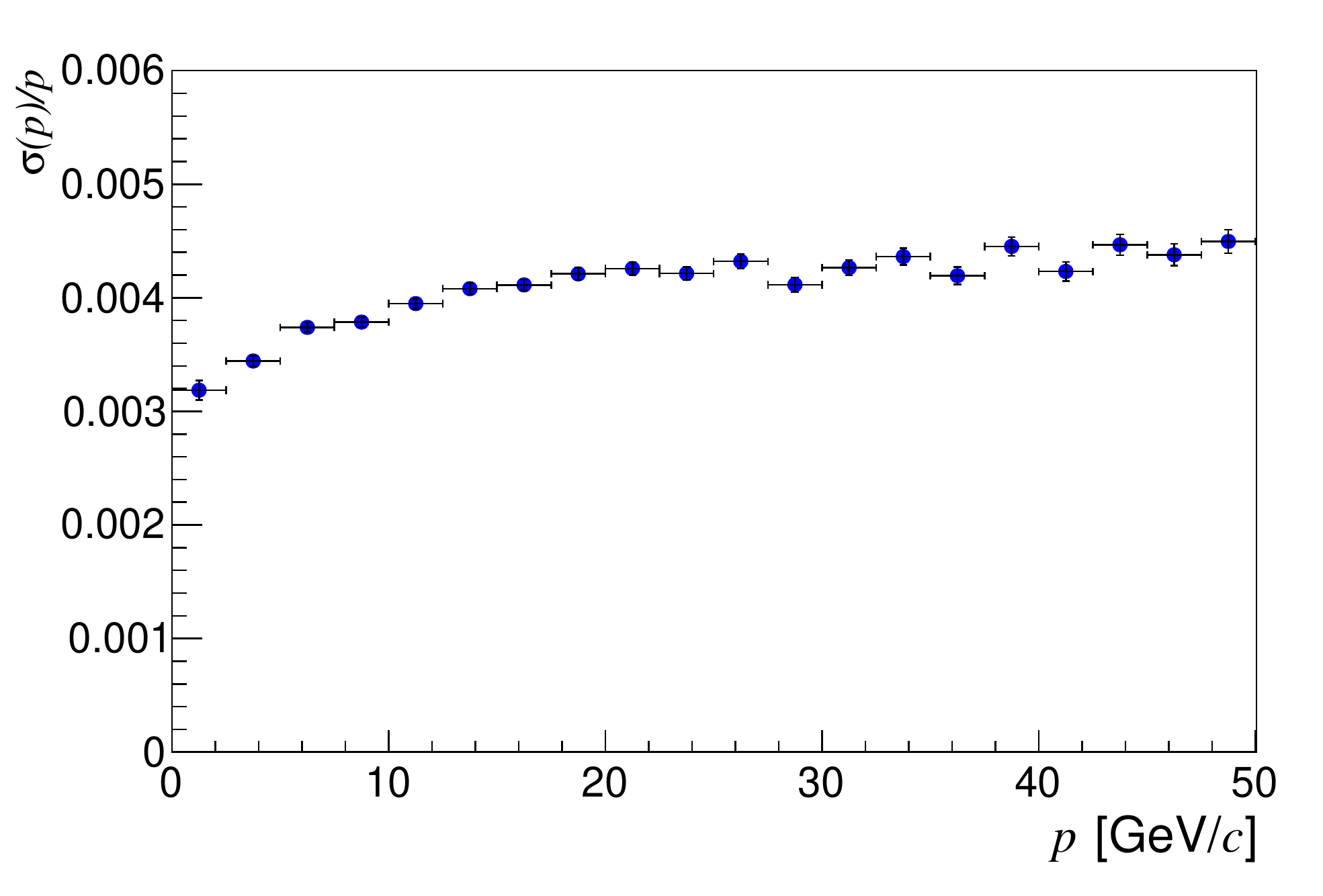}
    \caption{Left: Impact parameter resolution of child charged particles as a function of the inverse of the child's transverse momentum at 13 TeV. Right: Momentum resolution as a function of the child's momentum at 13 TeV. }
    \label{fig:performance}
\end{figure}

The detector simulation is an improved description of the fast simulation used in Ref.~\cite{Junior:2018odx}, adding multiple scattering and a more accurate description of the downstream tracking system. The simulated detector elements are the RF foil, the VeloPix (VP) stations, the Upstream Tracker, the Magnet, and the SciFi~\cite{LHCbCollaboration:2013bkh,LHCbCollaboration:2014tuj}, and are implemented in \texttt{python2}. No calorimetry or particle identification is simulated in our study. The particles generated by the procedure described in Section~\ref{sec:gen} are passed through detector elements and yield hits where appropriate. The set of hits is then passed to a track fit algorithm which calculates the track slopes at origin, the momentum, and a point in the early stage of the particle trajectory. The simulation also neglects occupancy effects and hit inefficiency. 
Benchmark figures of detector performance are shown in 
Figure~\ref{fig:performance}. The figures can be compared to the equivalent in the current full detector simulation (see Figures~30-33 in Ref.~\cite{LHCbCollaboration:2013bkh} and Figure~4.10 in Ref.~\cite{LHCbCollaboration:2014tuj}). It can be seen that the impact parameter resolution and momentum resolution are well reproduced at leading order, both in terms of detector resolution and in terms of multiple scattering.

\section{Evolution of detector efficiency with energy}
\label{sec:effs}

The accelerator energy changes the angular distribution of the produced particles and hence the probability for them to fall within the detector angular acceptance. In addition, for those particles with relatively long lifetimes, the different boost can slightly modify the fraction of those that decay within the acceptance of the vertex detector. To quantify the effect, in Table~\ref{tab:effs} we show the efficiency as a function of accelerator energy (relative to 13 TeV) for several benchmark decays from $b$-, $c$- and $s$-hadrons as well as for $\tau\rightarrow\mu\mu\mu$. It can be seen that there is a small gain of approximately $10\%$ in the acceptance for $b$-hadron decays when going to a high energy, while there is a small drop for $s$-hadron decays due to a higher fraction of them decaying outside of the acceptance of the vertex detector.

\begin{table}[]
\centering
\caption{Reconstruction efficiencies at 50 ($\varepsilon^{50}$) and 100 ($\varepsilon^{100}$) TeV normalized to those at $13$ TeV. The product of the efficiency ratio and the production cross-section ratio is also shown, since the event yield would be proportional to it. The efficiency ratios are calculated with and without a detachment requirement (impact parameter of the tracks greater than 100 microns), finding no difference at the level of the rounding digit. LL stands for kaons daughters reconstructed as long tracks (decays inside the VP acceptance), while DD stands for kaon daughters reconstructed as downstream tracks (decays outside the VP acceptance). The absolute reconstruction efficiency of $\Lambda_b^0\rightarrow K^+\pi^-\psi_{(DS)}$ decays is similar to that of $B_d^0\rightarrow\mu^+\mu^-$ as only two tracks are required.}
\begin{tabular}{c|c|c|c|c}
Decay     &  $\varepsilon^{50}/\varepsilon^{13}$ &$\varepsilon^{100}/\varepsilon^{13}$ & $\frac{\sigma_P^{50}\varepsilon^{50}}{\sigma_P^{13}\varepsilon^{13}}$ & $\frac{\sigma_P^{100}\varepsilon^{100}}{\sigma_P^{13}\varepsilon^{13}}$ \\ \hline
$B_d^0\rightarrow\mu^+\mu^-$     &  1.1 & 1.1 &  3.4 & 5.6 \\
$B_s^0\rightarrow J/\psi(\mu\mu)K^+K^-$ & 1.2 & 1.3 & 3.9 & 6.8 \\ 
$\Lambda_b^0\rightarrow K^+\pi^-\psi_{(DS)}(4340)$ &1.0 & 1.0 & 3.0 & 4.8 \\
$\Lambda_b^0\rightarrow K^+\pi^-\psi_{(DS)}(1460)$ &1.1 & 1.1 & 3.3 & 5.1 \\
$D^0\rightarrow K^+K^-$ & 1.00 & 0.99 & 2.2 & 3.07 \\
$D^0\rightarrow K_S^0 \text{(LL)} \pi^+ \pi^-$ & 0.97 & 0.94 & 2.13 & 2.91 \\
$D^0\rightarrow K_S^0 \text{(DD)} \pi^+ \pi^-$ & 1.08 & 1.07 & 2.38 & 3.32\\
$D^0\rightarrow \pi^+\pi^-\mu^+\mu^-$ & 1.08 & 1.10 & 2.38 & 3.41 \\
$K_S^0\rightarrow\mu^+\mu^- $ (LL)     &  0.89 & 0.84 &  1.4 & 1.7 \\
$K_S^0\rightarrow\mu^+\mu^- $ (DD)     &  1.0 & 1.0 &  1.6 & 2.0 \\
$\tau^-\rightarrow\mu^-\mu^+\mu^-$ & 1.03 & 1.07 & 2.58 & 4.07 \\
\end{tabular}
\label{tab:effs}
\end{table}

\section{Evolution of detector resolutions with energy}
\label{sec:res}
\begin{figure}[H]
    \centering
    \includegraphics[width=0.45\textwidth, page=1]{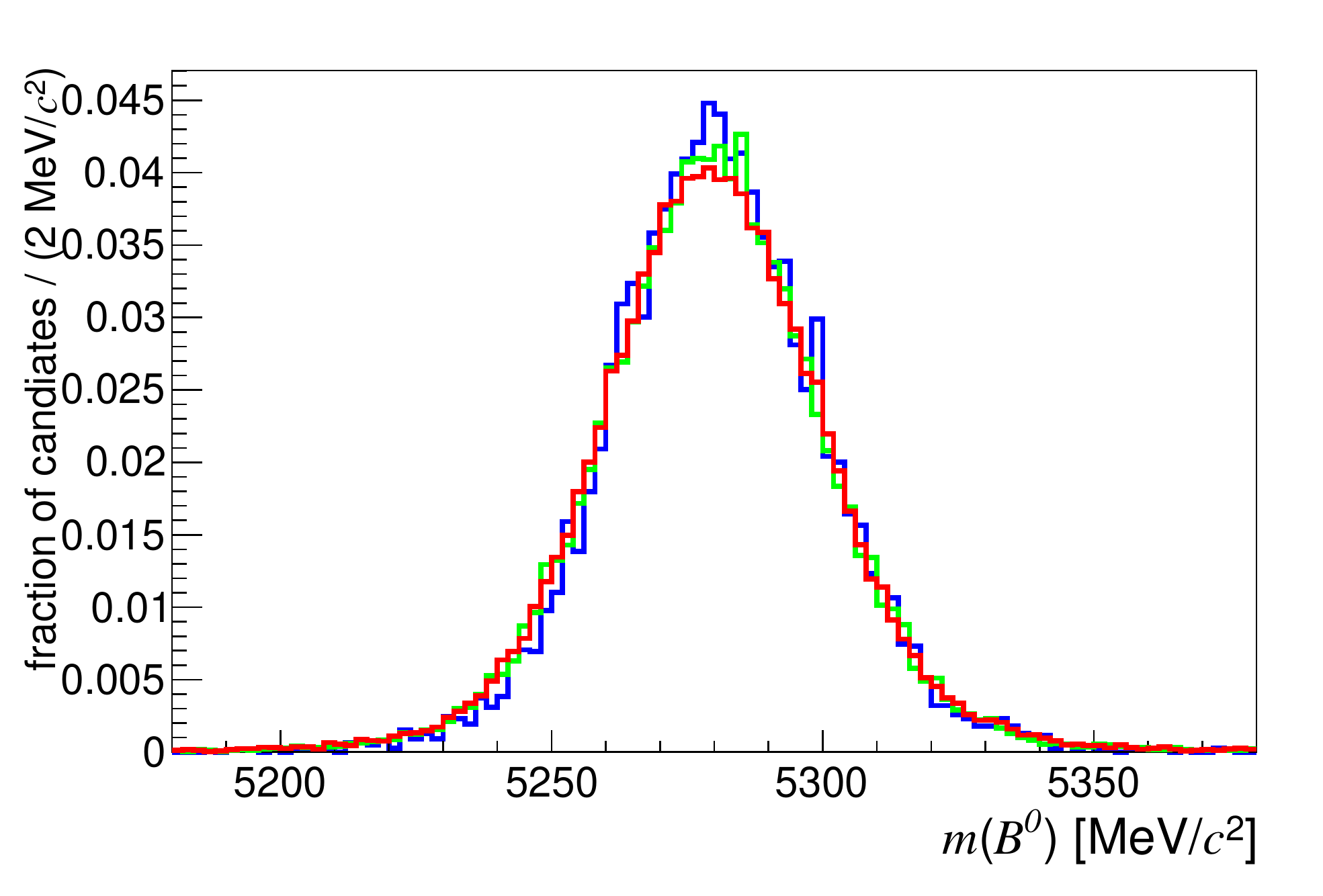}
    \includegraphics[width=0.45\textwidth, page=2]{MassAndTime.pdf}
    \caption{Left: Reconstructed mass peak for $B_d^0\rightarrow\mu^+\mu^-$ decays produced in our simulated collisions at different energies. A small ($13\%$) worsening of the mass resolution is observed at 100 TeV compared to LHC energies. Right: Propertime resolution for $B_s^0\rightarrow J/\psi(\mu^+\mu^-)K^+K^-$ decays produced in our simulated collisions at different energies. A few percent improvement in the propertime resolution is observed at 100 TeV compared to LHC energy. Blue: 13 TeV. Green: 50 TeV. Red: 100 TeV.}
    \label{fig:resolution}
\end{figure}

\begin{figure}[H]
    \centering
    \includegraphics[width=0.45\textwidth]{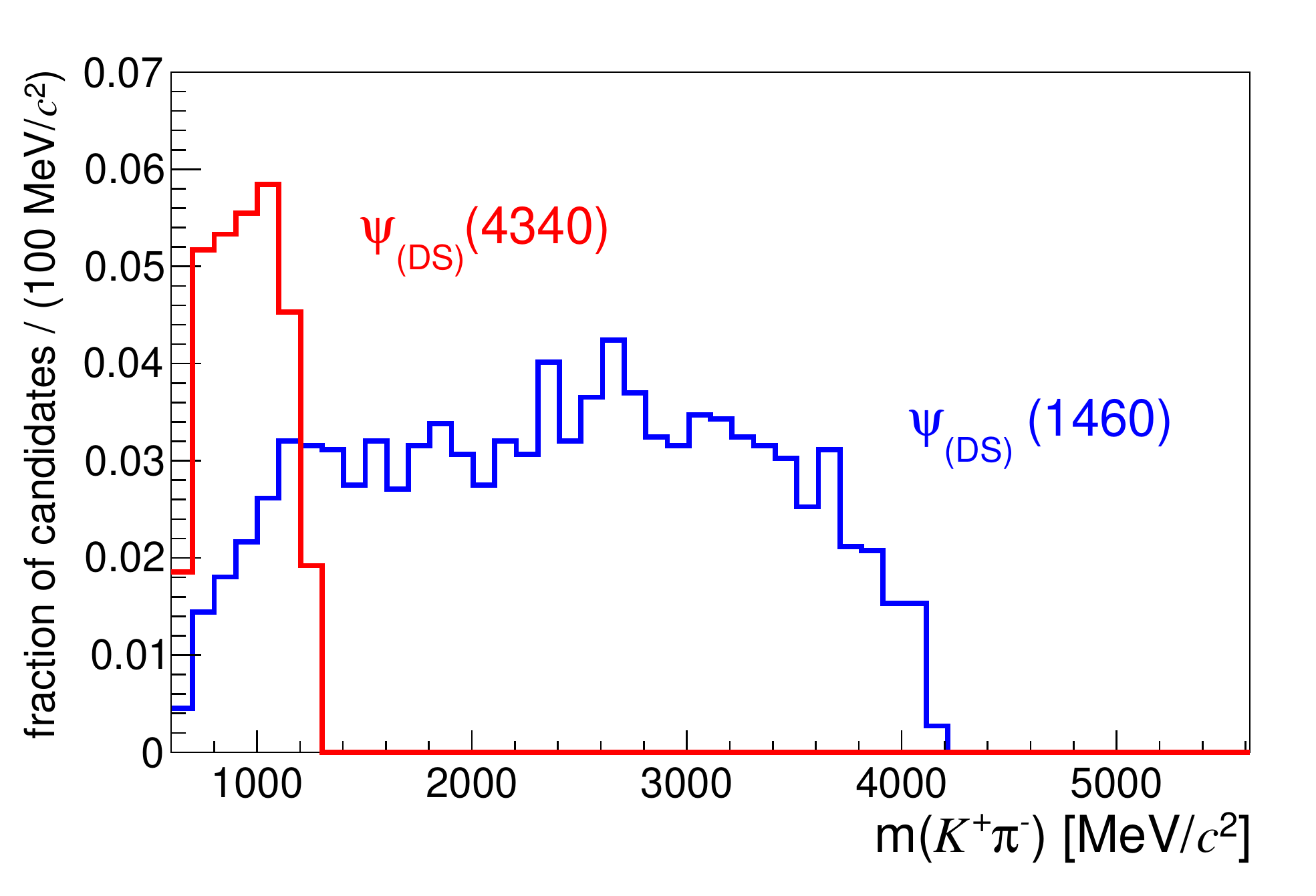}
    \includegraphics[width=0.45\textwidth]{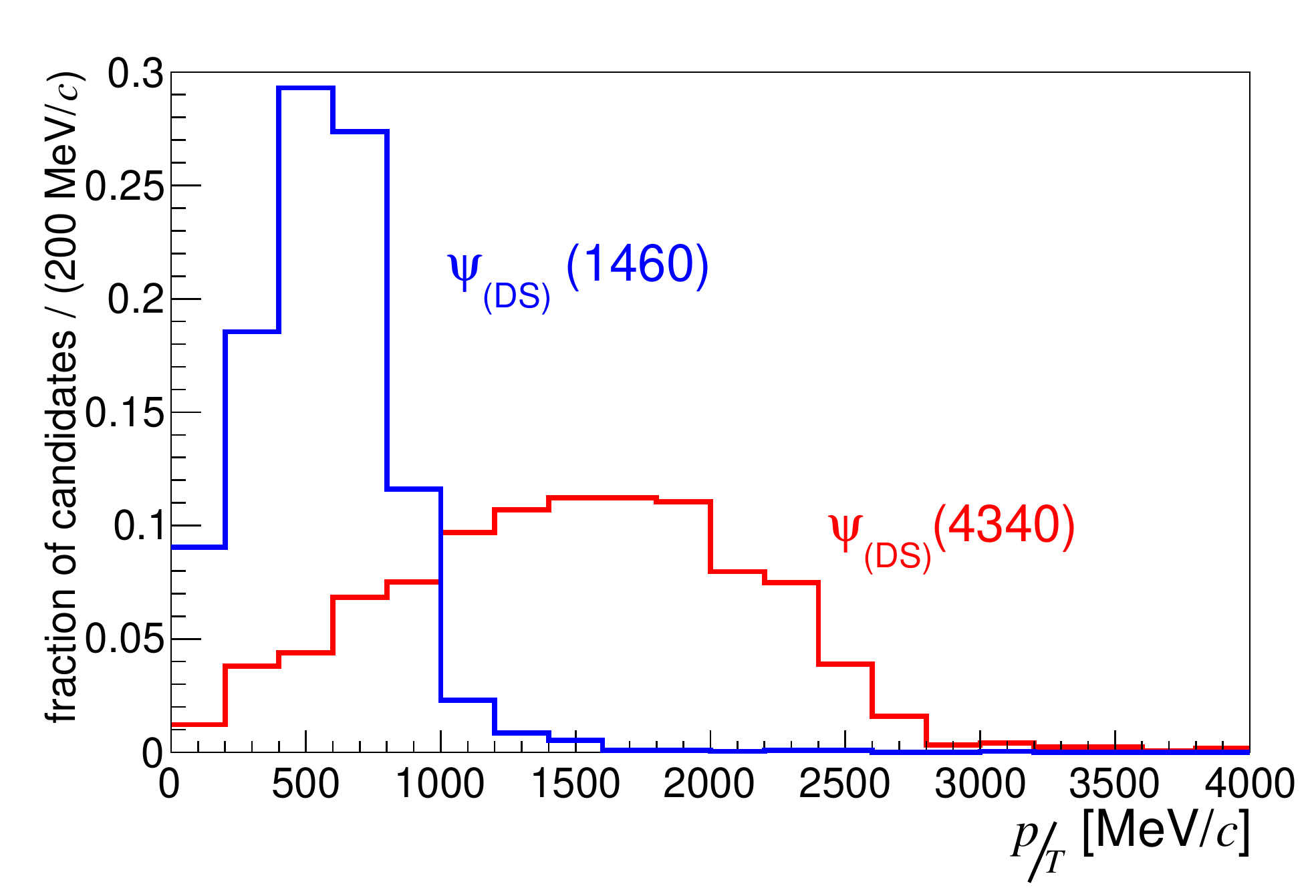}
    \caption{Left: Invariant mass distribution of $K^{+}\pi^{-}$ pairs from $\Lambda_b^0\rightarrow\psi_{(DS)}$ decays simulated at 100 TeV. A sharp kinematic end point at $M_{\Lambda_b^0} - M_{\psi_{(DS)}}$ is a distinctive signature of those decays. Right: Missing transverse momentum measured on those decays. None of these two distributions is significantly affected by the accelerator energy. The numbers in parentheses refer to the mass of the $\psi_{(DS)}$, and correspond to the maximum and minimum value allowed by other constraints. 
    }\label{fig:DM}
\end{figure}

The detector resolution also depends on the momentum of the particles and hence has in average some dependency on the collision energy. In particular, the momentum resolution worsens with energy, while on the other hand the impact parameter resolution improves with it. The invariant mass resolution is a function of both of them, and hence it may improve or worsen depending on the decay under study.
In Table~\ref{tab:res} we show the mass resolution as a function of accelerator energy (relative to 13 TeV) for several benchmark decays from $b$-, $c$- and $s$-hadrons as well as for \mbox{$\tau\rightarrow\mu\mu\mu$}.  In addition to the mass resolution, the propertime resolution is also important as it helps separating signal from background originating from the proton-proton collision, but, more importantly, it dilutes the amplitude of the measured $B_s^0-\overline{B}_s^0$ oscillation by the amount

\begin{equation}
D = e^{-\frac{1}{2}(\Delta m_s S_t)^2},
\end{equation}

\noindent which is equivalent to a signal efficiency multiplicative factor of $D$ for
measurements like $\phi_s$ in $B_s^0\rightarrow J/\psi K^+K^-$.
However, as seen in Table~\ref{tab:res}, the changes in the propertime
resolution are at the percent level and actually correspond to improved
resolution. This is because the vertex resolution improves as momentum
increases. Some distributions of mass and propertime resolution can be seen in Figure~\ref{fig:resolution}. In addition, flavor experiments also look for missing transverse energy signatures, to study decays with neutrinos, neutral pions, or search for dark sector particles. In Figure~\ref{fig:DM} we show some distinctive signatures for searches of dark sector particles. Larger boost and better decay vertex resolution lead to a better reconstruction of the missing transverse momentum. However, our studies showed no significant change in the distributions, due to the resolution being much smaller than the spread of the distribution.

\begin{table}[H]
\centering
\caption{Invariant mass (M) and propertime (t) resolutions at 50 ($S^{50}$ ) and 100 ($S^{100}$) TeV normalized to those at $13$ TeV. The muon pairs in $B_s^0\rightarrow J/\psi(\mu\mu)K^+K^-$  decays are constrained to have the $J/\psi$ mass. LL stands for kaon daughters reconstructed as long tracks (decays inside the VP acceptance), while DD stands for kaon daughters reconstructed as downstream tracks (decays outside the VP acceptance).}
\begin{tabular}{c|c|c|c|c}
Decay     &  $S_M^{50}/S_M^{13}$ &$S_M^{100}/S_M^{13}$ & $S_t^{50}/S_t^{13}$ & $S_t^{100}/S_t^{13}$ \\ \hline
$B_d^0\rightarrow\mu^+\mu^-$     &  1.1 & 1.1 &  1.0 & 1.0  \\\
$B_s^0\rightarrow J/\psi(\mu\mu)K^+K^-$ & 1.1 & 1.1 & 0.97 & 0.94 \\
$\Lambda_b^0\rightarrow K^+\pi^-\psi_{(DS)}(4340)$ & 1.0 & 1.1  & 0.95 & 0.91\\
$\Lambda_b^0\rightarrow K^+\pi^-\psi_{(DS)}(1460)$ &1.0 & 1.1 & 0.91 & 0.85\\
$D^0\rightarrow K^+K^-$ & 1.07 & 1.10 & 0.94 & 0.93 \\
$D^0\rightarrow K_S^0 \text{(LL)} \pi^+ \pi^-$ & 1.02 & 1.03 & 0.98 & 0.96 \\
$D^0\rightarrow K_S^0 \text{(DD)} \pi^+ \pi^-$ & 1.02 & 1.03 & 0.97 & 0.96 \\
$D^0\rightarrow \pi^+\pi^-\mu^+\mu^-$ & 1.04 & 1.06 & 0.95 & 0.95 \\
$K_S^0\rightarrow\mu^+\mu^-$ (LL)     &  1.0 & 1.0 & 0.94 & 0.91 \\
$K_S^0\rightarrow\mu^+\mu^-$ (DD)     &  1.0 & 1.1 & 0.96 & 0.94 \\
$\tau^-\rightarrow\mu^-\mu^+\mu^-$ & 0.94 & 1.00 & 0.93 & 0.95 \\
\end{tabular}
\label{tab:res}
\end{table}

\section{Conclusions}
\label{sec:conclusions}
We presented the results of a simulation of proton-proton collisions at 13, 50 and 100 TeV recorded by a simplified implementation of an LHCb-like detector. We obtain a significant increase of the expected yield per $fb^{-1}$ with the center-of-mass-energy for several interesting flavor decays, at least ignoring occupancy effects. The detector resolutions are not significantly affected by the different boosts obtained in each center-of-mass energy.

\section{Acknowledgements}
This work has received financial support from Xunta de Galicia (Centro singular de investigaci\'on de Galicia accreditation 2019-2022), by European Union ERDF, and by  the “Mar\'ia  de Maeztu”  Units  of  Excellence program  MDM-2016-0692  and  the Spanish Research State Agency. Veronika Chobanova is supported by MCINN (Spain) through the Juan de la Cierva-incorporaci\'{o}n program with grant IJCI-2017-32371. We would like to thank M. Escudero and G. Alonso-\'Alvarez for discussions on dark sector signatures.
\clearpage
\bibliographystyle{unsrt}
\bibliography{main.bib}

\end{document}